# Matrix Elements of the Singlet Axial Current in the Proton

Rajan Gupta

*T-8, MS-B285, Los Alamos National Laboratory, Los Alamos, NM 87545*

Jeffrey E. Mandula

*Department of Energy, Division of High Energy Physics, Washington, DC 20585*

We present a method to estimate the matrix element of the singlet axial current within a polarized proton state using lattice QCD. The method relies on using the Adler-Bell-Jackiw anomaly and gives the desired result in the chiral limit. We show that this method fails in the quenched approximation. For heavy quarks one does not expect much difference between simulations including dynamical quarks and those done in the quenched approximation. For that reason we explore numerical methods on an existing set of quenched configurations. The data obtained in this exploratory study show a poor statistical signal.

26 Aug 1994

## 1. Introduction

Four years ago, the European Muon Collaboration (EMC) measured the spin structure of protons using deep inelastic muon scattering from protons. Their principal result, which stirred a great deal of interest, was [1]

$$\int_0^1 g_1^P(x)dx = 0.126 \pm 0.010 \pm 0.015 \tag{1.1}$$

The quoted systematic error of 0.015 includes an estimate of the uncertainty due to the need to extrapolate the measured value of the structure functions to the regions above Bjorken $x \sim 0.7$ and below $x \sim 0.02$. The first moment of $g_1$ is related to the spin structure of the proton through

$$2\int_0^1 g_1^P(x,Q^2)dx = (1 - \frac{\alpha_s(Q^2)}{\pi})\frac{1}{9}[4\Delta u(Q^2) + \Delta d(Q^2) + \Delta s(Q^2)] + \ldots \tag{1.2}$$

where corrections due to heavy quarks and higher twist effects have been ignored. The quantity $\Delta q$ is the forward matrix-element of the quark axial current inside a proton,

$$s^\mu \Delta q(\mu^2) = \langle \vec{p}, s | \bar{q} i \gamma_\mu \gamma_5 q \Big|_{\mu^2} | \vec{p}, s \rangle . \tag{1.3}$$

Here $s_\mu$ is the covariant spin vector of the proton and $\mu^2$ is the renormalization point for the axial current. Since the singlet axial current picks up a small anomalous dimension only at two loops [2] we, therefore, ignore this dependence on the renormalization point in the following discussion. The EMC result (1.1) gives one linear combination of $\Delta u$, $\Delta d$ and $\Delta s$. A second relation is obtained from the neutron $\beta$-decay

$$(\Delta u - \Delta d) = g_A \tag{1.4}$$

where $g_A = 1.2573 \pm 0.0028$ is the axial vector coupling [3]. A third relation between the quark spin fractions can only be derived assuming SU(3) flavor symmetry. It is

$$(\Delta u + \Delta d - 2\Delta s) = 3F_A - D_A \tag{1.5}$$

where $F_A$ and $D_A$ are reduced matrix elements of the axial current. From an analysis of the strangeness changing semi-leptonic decays of hyperons, Jaffe and Manohar obtain [4]

$$F_A = 0.47 \pm 0.04; \qquad D_A = 0.81 \pm 0.03. \tag{1.6}$$

Combining Eqs. (1.1) - (1.6) they concluded that

$$\Delta u = +0.74 \pm 0.10$$
$$\Delta d = -0.54 \pm 0.10 \qquad (1.7)$$
$$\Delta s = -0.20 \pm 0.11$$

and that the total axial charge of the nucleon is

$$\Delta \Sigma = \Delta u + \Delta d + \Delta s = 0.01 \pm 0.29 \qquad (1.8)$$

at $Q^2 = Q^2_{EMC}$.

Since in the parton model $\Delta \Sigma$ gives the fraction of spin carried by the quarks, the import of Eq. (1.8) is that the quarks contribute only a small fraction of the proton's spin. While this interpretation cannot strictly be derived from QCD, both this and the second consequence of Eq. (1.7) that there is a significant contribution from the strange quarks in the sea, are certainly challenges to our understanding of the structure of hadrons.

The experimental situation has changed considerably in the last year. The SLAC E-142 collaboration has extracted $\Delta \Sigma$ and $\Delta s$ from polarized electron scattering from a polarized $^3He$ target [5]. Their results are

$$\Delta \Sigma = +0.57 \pm 0.11$$
$$\Delta s = -0.01 \pm 0.06 \ . \qquad (1.9)$$

More recently, the SMC collaboration [6] have reported the values

$$\Delta \Sigma = +0.36 \pm 0.21$$
$$\Delta s = -0.07 \pm 0.06 \qquad (1.10)$$

from polarized muon scattering off polarized deuterium. A global analysis of the data has also been made by Ellis and Karliner [7] including higher twist effects to compensate for the different values of $Q^2$ at which the three experiments are carried out. Using the values

$$F_A = 0.46 \pm 0.01; \qquad D_A = 0.79 \pm 0.01 \ . \qquad (1.11)$$

taken from Ref. [8] they report

$$\Delta u = +0.80 \pm 0.04$$
$$\Delta d = -0.46 \pm 0.04 \qquad (1.12)$$
$$\Delta s = -0.13 \pm 0.04$$

and
$$\Delta\Sigma = \Delta u + \Delta d + \Delta s = 0.22 \pm 0.10 \ . \tag{1.13}$$

The difference between these estimates of $\Delta\Sigma$ reflects to some extent the systematic error in the analysis due to the uncertainty in the contribution of small $x$ region to $g_1$, the uncertainty in including higher twist effects, and the uncertainty in the values of $F_A$ and $D_A$. Given the large spread in the results it is clear that the ability to calculate the matrix element of the singlet axial current in a proton will improve our understanding of this issue considerably.

Numerical simulations of lattice QCD, in principle, provide an exact non-perturbative approach to calculating the matrix elements of the singlet axial current inside the proton. In Section 2 we present a method to extract the singlet matrix element using the Adler-Bell-Jackiw anomaly. In Section 3 we describe the problems with using this method in the quenched approximation and show that the matrix element is expected to diverge in the chiral limit. We discuss the lattice implementation in Section 4 and the subleties of calculating the topological charge density in Section 5. Assuming that for quarks with mass $m_q \sim m_s$ the effect of vacuum polarization is small, we attempt an exploratory calculation using the quenched approximation to explore numerical methods for a possible signal. The lattice parameters used are given in section 6 and the analysis of the data, which show a poor statistical signal, is presented in section 7. Our final conclusions are contained in Section 8.

## 2. Relation between the axial current and the $U(1)$ anomaly

There are two possible Wick contractions that contribute to the matrix elements of the flavor singlet axial current inside a proton. These are shown in Fig. 1 and labeled as the valence and disconnected graphs. Unfortunately, of these two the one measuring the correlation of the disconnected loop with the proton has a very poor signal. A preliminary lattice estimate found statistical errors nearly twice the measured value of the strange quark spin fraction [9]. Since this amplitude is not *a priori* tiny, contributes for each quark flavor, and is responsible for all of $\langle \vec{p}, s | \bar{s} i \gamma_\mu \gamma_5 s | \vec{p}, s \rangle$, we exploit an alternate method which is especially suitable to extract the total quark axial charge of the nucleon [10]. In the chiral limit the anomaly condition relates the divergence of the flavor singlet axial current, $A_\mu$, to the purely gluonic operator $F\tilde{F}$

$$\partial_\mu A_\mu = N_f \frac{\alpha_s}{2\pi} \text{Tr} F_{\mu\nu} \tilde{F}_{\mu\nu}, \qquad \tilde{F}_{\mu\nu} = \frac{1}{2} \epsilon_{\mu\nu\rho\sigma} F_{\rho\sigma}, \tag{2.1}$$

where $N_f = 3$ is the number of active flavors. Taking the matrix elements of these operators inside a proton state and transforming to momentum space gives the relation

$$\Delta\Sigma = \langle \vec{p}, s | A_\mu | \vec{p}, s \rangle s_\mu = N_f \frac{\alpha_s}{2\pi} \lim_{\vec{q} \to 0} \frac{-i|\vec{s}|}{\vec{q} \cdot \vec{s}} \langle \vec{p}', s | \text{Tr} F_{\mu\nu} \tilde{F}_{\mu\nu}(\vec{q}) | \vec{p}, s \rangle \qquad (2.2)$$

where $\vec{q} = \vec{p} - \vec{p}'$ and the two independent directions of proton spin are taken to be $\vec{s}/|s| = \pm \vec{q}/|\vec{q}|$. Thus, operationally to get the final result we have to calculate the matrix elements between states with different momenta, divide by the momentum transfer, and then extrapolate to zero momentum transfer. The strong coupling constant $\alpha_s$ should be evaluated at the same $Q^2$ at which the experimental result has been extracted even though the anomalous dimension of $\alpha_s/2\pi \text{Tr} F\tilde{F}$ is zero. The reason is that $\alpha_s/2\pi \text{Tr} F\tilde{F}$ and $A_\mu$ mix at 1-loop [2], have the same $Q^2$ evolution and thus have to be defined at the same renormalization point.

Most numerical calculations are at present done in the quenched approximation. Therefore, in the next section we analyze the problems that arise in that case. We find that due to the vanishing of the $\eta'$ mass as $m_q \to 0$ one cannot calculate the spin asymmetry from the matrix element of $F\tilde{F}$ alone and the proposed method cannot be used to get the desired result.

## 3. The quenched approximation

One *a priori* expects that vacuum polarization effects play an important role in the matrix elements of the singlet axial current due to the different behavior of $\eta'$ in the quenched and full theories. In particular, since the $\eta'$ becomes massless in the chiral limit of the quenched theory, the result may be singular. We show that this is indeed true and that in the quenched approximation Eq. (2.2) cannot be used to extract an estimate for the desired matrix element.

We start with the Fourier transform of the anomaly relation for the matrix element of the flavor singlet axial current

$$iq_\mu \langle \vec{p}', s | A_\mu(q) | \vec{p}, s \rangle = 2m_q \langle \vec{p}', s | P | \vec{p}, s \rangle + N_f \frac{\alpha_s}{2\pi} \langle \vec{p}', s | \text{Tr} F\tilde{F} | \vec{p}, s \rangle \qquad (3.1)$$

where $P$ is the SU(3) singlet pseudoscalar density operator. The first two terms have both connected and disconnected diagrams while the $F\tilde{F}$ term only contributes through a disconnected diagram. These terms have single and double poles at $q^2 = m_{\eta'}^2$ due to the $\eta'$ propagator [11]. We illustrate these in Figs. 2a-e. It turns out that in the $q^2 \to 0$ and $m_q \to 0$ limit the desired matrix element is finite even in the quenched theory; nevertheless it cannot be calculated from $\langle \vec{p}', s | \operatorname{Tr} F\tilde{F} | \vec{p}, s \rangle$ alone.

We parameterize each of the three matrix elements present in Eq. (3.1) in terms of form-factors as

$$\langle \vec{p}', s | A_\mu(q) | \vec{p}, s \rangle = \overline{u} i \gamma_\mu \gamma_5 u G_1^A - i q_\mu \overline{u} \gamma_5 u G_2^A,$$
$$\langle \vec{p}', s | P | \vec{p}, s \rangle = \overline{u} \gamma_5 u G^P, \quad (3.2)$$
$$\langle \vec{p}', s | \operatorname{Tr} F\tilde{F} | \vec{p}, s \rangle = \overline{u} \gamma_5 u G^F.$$

Then, with the spinor normalization

$$u_{p,s} \overline{u}_{p,s} = \frac{(\not{p} + M_P)}{2 M_P} \frac{1 + i \gamma_5 \not{s}}{2}, \quad (3.3)$$

one has the relation $G_1^A(q^2 = 0) = \Delta \Sigma$. The singularities in these form factors with respect to $q^2$ and due to the $\eta'$ propagators are

$$G_2^A(q^2) = \frac{a_2}{(q^2 - m_{\eta'}^2)^2} + \frac{a_1}{(q^2 - m_{\eta'}^2)} + \tilde{G}_2,$$
$$G^P(q^2) = \frac{p_2}{(q^2 - m_{\eta'}^2)^2} + \frac{p_1}{(q^2 - m_{\eta'}^2)} + \tilde{P}, \quad (3.4)$$
$$G^F(q^2) = \frac{f_1}{(q^2 - m_{\eta'}^2)} + \tilde{F}.$$

Note that the $\eta'$ can couple to the proton through pseudoscalar and axial couplings. However, both of these are proportional to $q_\mu \overline{u} \gamma_5 u$ for on-shell matrix elements. The form factor $G_1^A$ gets contributions from contact terms and the exchange of $1^+$ states, but not from $\eta'$ propagating as an intermediate state. Thus, only $G_2$ contains $\eta'$ pole terms. Equating the single and double pole terms gives two relations

$$m_{\eta'}^2 a_2 = 2 m_q p_2$$
$$m_{\eta'}^2 a_1 = \frac{-2 m_q p_2}{m_{\eta'}^2} + 2 m_q p_1 + N_f \frac{\alpha_s}{2\pi} f_1 \quad (3.5)$$

between the residues. Now using Eqs. (3.1) and (3.2) we can write

$$2 M_P G_1^A(q^2) = -q^2 G_2^A(q^2) + 2 m_q G^P(q^2) + N_f \frac{\alpha_s}{2\pi} G^F(q^2) \quad (3.6)$$

which has the limiting behavior

$$\lim_{q^2 \to 0} 2 M_P G_1^A(q^2) = -a_1 + 2 m_q \tilde{P} + N_f \frac{\alpha_s}{2\pi} \tilde{F},$$
$$\lim_{m_q \to 0} 2 M_P G_1^A(q^2) = -a_1 - q^2 \tilde{G}_2 + N_f \frac{\alpha_s}{2\pi} \tilde{F}, \quad (3.7)$$

showing that $G_1^A$ is finite in the chiral and $q^2 \to 0$ limit. In the lattice calculation we first take $q^2 \to 0$ and then extrapolate to the chiral limit. In this double limit one gets the result

$$2M_P G_1^A(q^2 = 0) = -a_1 + N_f \tfrac{\alpha_s}{2\pi} \tilde{F},$$
$$= \frac{2m_q}{m_{\eta'}^2}\left(\frac{p_2}{m_{\eta'}^2} - p_1\right) + N_f \tfrac{\alpha_s}{2\pi}\left(\tilde{F} - \frac{f_1}{m_{\eta'}^2}\right). \tag{3.8}$$

The second expression is obtained using Eq. (3.5). Using the anomaly relation (2.2) to calculate the spin asymmetry gives only the term proportional to $N_f \tfrac{\alpha_s}{2\pi}$. This diverges in the chiral limit and there is no obvious way of extracting the physical answer from it. Thus, to get the result in the quenched approximation, the best option is to calculate the connected and disconnected diagrams for the axial current itself.

In the full theory, there are no double poles as the diagrams in Figs. 2b-d, along with all diagrams related by the insertion of vacuum polarization bubbles, are part of the single $\eta'$ propagator shown in Fig. 2f. Therefore, keeping only single pole terms in Eq. (3.4) and repeating the above analysis, one finds the relation between the residues to be

$$m_{\eta'}^2 a_1 = 2m_q p_1 + N_f \tfrac{\alpha_s}{2\pi} f_1. \tag{3.9}$$

In the double limit $q^2 \to 0$, $m_q \to 0$ one gets

$$2M_P G_1^A(q^2 = 0) = -a_1 + N_f \tfrac{\alpha_s}{2\pi} \tilde{F},$$
$$= N_f \tfrac{\alpha_s}{2\pi}\left(\tilde{F} - \frac{f_1}{m_{\eta'}^2}\right), \tag{3.10}$$

which is precisely the anomaly relation (2.2).

To summarize, our analysis exposes a very serious shortcoming of the quenched approximation. The matrix element one gets using Eq. (2.2) is expected to diverge in the chiral limit. The most one can hope for is that for quarks of mass $\geq m_s$ there is little effect of quark loops as one is far from the chiral limit. In this region one can use quenched lattices to develop numerical tricks to improve the signal that will be relevant for calculations in the full theory. Thus our present numerical work should be regarded as an exploration of methods.

## 4. Lattice implementation

On the lattice we calculate the following correlation function

$$\text{Re}\left(\frac{-i}{|\vec{q}|}\langle N_{\vec{p}'}(t)(P_+ - P_-)\text{Tr}F_{\mu\nu}\tilde{F}_{\mu\nu}(\tau, \vec{q})N_{\vec{p}}(0)\rangle\right) \qquad (4.1)$$

where $N_{\vec{p}}(t)$ is a creation/annihilation operator for the proton at time $t$ and momentum $\vec{p}$, $P_\pm$ are the spin projection operators, and in Euclidian space $P_+ - P_- = i\slashed{s}\gamma_5$ projects out the spin asymmetry. In order to extract the desired matrix element we have to remove the amplitude for creating and annihilating the proton and the exponential damping in time from the above correlator. In addition it is necessary to choose $t \gg \tau \gg 0$ so that only the lowest proton state contributes. To achieve this we calculate the following ratio of correlators

$$\mathcal{R}(t, \tau) = \frac{\text{Re}\left(\frac{-i}{|\vec{q}|}\text{Tr}\langle N_{\vec{p}'}(t)(P_+ - P_-)F_{\mu\nu}\tilde{F}_{\mu\nu}(\tau, q)N_{\vec{p}}(0)\rangle\right)}{\text{Re}\left(\text{Tr}\langle N_{\vec{p}}(t)\frac{(1+\gamma_4)}{2}N_{\vec{p}}(0)\rangle\right)} \qquad (4.2)$$

It is easy to see that in this ratio the normalization of the proton field, including the Wuppertal smearing factors, cancel. In addition, the exponential decay factor $e^{-E_N t}$ is cancelled if the magnitude of the momentum is unchanged between the inital and final proton state, i.e. $|\vec{p}| = |\vec{p}'|$.

We have calculated this ratio for three classes of momentum transfer: $(1, 0, 0) \to (0, 1, 0)$, $(1, 0, 0) \to (-1, 0, 0)$, and $(2, 0, 0) \to (-2, 0, 0)$. In each case the results are averaged over all permutations of the axes, and over forward and backward propagation on a given background configuration. Statistical errors are then computed by a single elimination jackknife method regarding each of the 35 data points as statistically independent.

The cases $(1, 0, 0) \to (-1, 0, 0)$ and $(2, 0, 0) \to (-2, 0, 0)$ are simple to analyze. We choose the two independent directions of proton spin to be $\vec{s}/|s| = \pm\vec{p}/|\vec{p}|$. The case $(1, 0, 0) \to (0, 1, 0)$ has the advantage that, under the restriction that $|\vec{p}| = |\vec{p}'|$, the momentum transfer is the smallest possible. In this case we take the spin projection to be along the direction of momentum transfer $\vec{q} = (\vec{p} - \vec{p}')$.

## 5. Topological density

The calculation of the topological density $\mathcal{Q}(x) \equiv \text{Tr} F_{\mu\nu} \tilde{F}_{\mu\nu}$ on the lattice has not proven to be straightforward. In the continuum limit, we require that the topological charge, given by $Q = -1/16\pi^2 \int_x \mathcal{Q}(x)$, take on integer values and be related to fermionic zero modes, *i.e.* $Q$ is the difference between the number of left and right handed zero modes. These properties hold on the lattice near the continuum limit where the lattice gauge fields are smooth. The problem with present lattice calculations ($\beta \approx 6.0$) is that the gauge fields are not smooth. As a result there exist lattice artifacts, called dislocations, that are caused by ultraviolet fluctuations, and these affect the calculation of $\mathcal{Q}$ and need to be removed before one can get sensible results. Also, on the lattice $\mathcal{Q}(x)$ mixes with the other dimension four pseudo-scalar operator $m\overline{\psi}\gamma_5\psi$ and only a linear combination is equal to the divergence of the axial current. Our method of using the Adler-Bell-Jackiw relation circumvents this problem, on the other hand the results are only valid in the chiral limit.

We have used the naive lattice transcription of $\text{Tr} F_{\mu\nu}\tilde{F}_{\mu\nu}$ where the gluon field strength $F_{\mu\nu}(x)$ is defined as the traceless anti-hermitian part of the plaquette. To improve the estimate with respect to $\mathcal{O}(a)$ corrections we use the symmetrized average of the four plaquettes in the $[\mu,\nu]$ plane [12]. While this field theoretic definition (which we shall call the plaquette method) has the correct continuum limit, it is well known that $Q$ does not take on integer values [13] at finite lattice spacing and it is believed that the contribution of lattice artifacts may be large. We discuss this point and operator renormalization in some detail below.

A number of numerical methods have been proposed to smooth out the gauge fields without changing the topological character of the configuration. These methods, which go under the name of "cooling" and "smearing", have been compared in, for example, Ref. [14] but it is still not clear whether the smoothing process washes out only the lattice artifacts or in addition some physical features of $\mathcal{Q}(x)$. It is therefore important to calculate $\mathcal{Q}(x)$ a number of different ways and check whether the result for the matrix element is the same.

The success of all the plaquette methods relies on the validity of a decomposition of the form [15]

$$\mathcal{Q}_L(x) = \big(Z(\beta) + \zeta(x)\big)\mathcal{Q}(x) + \eta(x) \tag{5.1}$$

where $\zeta(x)$ and $\eta(x)$ are random variables with zero mean, and $Z(\beta)$ is the renormalization constant. Numerical data for the O(3) non-linear sigma model in two dimensions show that such a decomposition is justified and $Z(\beta)$ agrees with the 2-loop perturbative result [15]. We shall assume that this is true also for 4-D gauge theories. The problem then reduces to one of getting adequate statistics, *i.e.* averaging over a sufficient number of gauge configurations such that the correlations of $\zeta(x)\mathcal{Q}(x)$ and $\eta(x)$ with the proton correlator are zero. This assumption can be tested by comparing results using different methods of calculating $\mathcal{Q}(x)$.

Assuming that the decomposition given in Eq. (5.1) works, one still needs the multiplicative renormalization factor $Z(\beta)$. This has been calculated in perturbation theory and the result of Ref. [16] is

$$Z(\beta) = 1 + \frac{Z_1}{\beta} + \frac{Z_2}{\beta^2} \tag{5.2}$$

where $Z_1 = -5.4508$ and $Z_2 \sim 11.1$. (In the 2-loop part $Z_2$ only the dominant graphs have been evaluated). Thus $Z(\beta = 6.0) \approx 0.4$. The renormalization factor was also estimated using a cooling procedure [16], which gave $Z(\beta = 6.0) \approx 0.3 \pm 0.07$.

An estimate of $Z(\beta)$ can also be obtained using the mean-field method of Lepage and Mackenzie [17]

$$Z = U_0^8 = U_{plaquette}^2 \approx 0.35 \tag{5.3}$$

where $U_0^4$ is the expectation value of the plaquette. The three estimates are consistent and show that the renormalization effects are very large. In this study we shall present results using the mean-field value given in Eq. (5.3).

An alternate approach to the plaquette method is to use Lüscher's geometrical definition of $\mathcal{Q}(x)$. This, by construction, guarantees that $Q$ takes on integer values and has the correct continuum limit. However, it has been shown that even in this method dislocations contribute to $\mathcal{Q}(x)$ for the simple Wilson action [18]. The suggested cure is to use an improved gauge action which increases the action of dislocations so that they are supressed for $\beta \geq 6.0$ and vanish in the continuum limit. Recently, Schierholz has argued that at corresponding values of the lattice scale, dislocations are also suppressed in simulations with light dynamical fermions [19]. We are in the process of developing codes to calculate $\mathcal{Q}(x)$ using Lüscher's method so as to compare the results with the plaquette method (with and without using smeared or cooled lattices).

To summarize, the plaquette method for calculating $\mathcal{Q}(x)$ is computationally much faster, however it does assume that Eq. (5.1) is valid and that $Z(\beta)$ can be estimated reliably. Since all lattice methods for calculating $\mathcal{Q}(x)$ have potential problems and the statistical signal is poor, we consider it worthwhile to compare results for the matrix element with different definitions of $\mathcal{Q}(x)$ on the same set of lattices. The use of the plaquette method should be considered as the first step towards this goal.

## 6. Details of the lattices and quark propagators

The calculation uses 35 quenched gauge configurations of size $16^3 \times 40$ at $\beta = 6.0$. The Wilson action quark propagators were calculated on doubled lattices ($16^3 \times 40 \rightarrow 16^3 \times 80$) using Wuppertal sources. The quark masses used are $\kappa = 0.154$ and $0.155$, corresponding to pions of mass 700 and 560 MeV respectively. More complete details of lattice generation and propagator inversion are given in Ref. [20]. These parameter values are far from the real world, and as discussed in Section 3 we do not expect realistic results in the quenched approximation. Nevertheless, quenched lattices form a good laboratory to test numerical methods, so the goal of our calculation is to explore techniques to enhance the signal. Unfortunately, our conclusion is that an improvement in statistics and numerical methods will be required to get a reliable signal in eventual simulations with dynamical quarks.

## 7. Analysis

There are three quantities which control the quality of the final signal. We discuss each of these below.

### 7.1. The operator $\langle F\tilde{F} \rangle$:

The expectation value of $\mathcal{Q}(t,q)$ should be zero on each time slice and for each value of momentum transfer. Our data satisfy this condition within the error estimate for all three cases of momentum transfer. The data for the best case, $\vec{p} = (1,1,0)$, is shown in Fig. 3 and we find that the signal is very noisy and that the correlations between successive time-slices are large.

### 7.2. Proton Correlators

At large time separations the proton correlator should be saturated by the lowest state. The data for $\vec{p} = (1,0,0)$ are shown in Fig. 4. We find a plateau in the effective mass for $\vec{p} = (0,0,0)$ and $\vec{p} = (1,0,0)$ and a marginal signal for $\vec{p} = (2,0,0)$. The saturation ($> 95\%$) by the lowest state sets in for $t > 8$ in each case as examplified in Fig. 4. This implies that in order to get a reliable estimate for the matrix element, the insertion of $F\tilde{F}$ should be at least 8 time slices from both the proton source and sink. It is hoped that some of the effects of the higher states in the proton correlator cancel in the ratio defined in Eq. (4.2) so that one can extract reliable estimates of the ground state matrix element from slightly smaller separations. Unfortunately, as we discuss below the signal in our data is not good enough to draw any reliable conclusions about the stability of the results.

### 7.3. Correlation of $F\tilde{F}$ with the Proton

The correlation between $F\tilde{F}$ and the proton occurs as a result of gluon exchange in the $t$-channel. This is shown schematically in Fig. 2e. The analogue of this in the $s$ channel is the off-diagonal matrix element for the production of a pair of nucleons by a glueball. Based on glueball calculations one expects that to get a good signal will require a large number of configurations (many more than the 35 configurations used in this study). However the situation may not be this bad. The correlation of $F\tilde{F}$ with the proton comes mainly from contact terms, i.e. correlated fluctuations of the same link variables that enter into $F\tilde{F}$ and also into the Dirac equation for the quark propagators. This suggests that it is the short distance correlations which give the dominant contribution. So the signal one hopes for may have more in common with the computation of glueball propagators at zero and small separations than that at large separations as needed for extracting masses. It is this physical picture that gives us hope of a reasonable signal in the present calculation.

The data for the three different cases of momentum transfer, $(1,0,0) \to (-1,0,0)$, $(2,0,0) \to (-2,0,0)$ and $(1,0,0) \to (0,1,0)$, are shown in Tables 1 for the heavier mass $\kappa = 0.154$. All numbers should be multiplied by $10^{-2}\alpha_s$ to get $\Delta\Sigma$. The data are shown as a function of the time slice at which the proton is annihilated ($T_P$) and at which the operator ($T_{F\tilde{F}}$) is inserted. The proton is created at time-slice $T = 0$. We find that the statistical errors in the results are such that no signal is evident, and that these data are statistically indistinguishable from zero.

Qualitatively, we find as expected that the statistical errors are best for the case $(1,0,0) \to (0,1,0)$, somewhat larger for $(1,0,0) \to (-1,0,0)$ and still larger for the case $(2,0,0) \to (-2,0,0)$. Also the errors increase with $T_P$ and the maximum separation we can probe is $T_P = 9$, significantly short of the requirement that the insertion of $F\tilde{F}$ should be at least 8 time-slices from either end. In view of this lack of signal, some future possibilities to improve the signal are to work on a larger lattice in order to reduce the value of $|q|$, to use alternate definitions of $\mathcal{Q}(x)$, to find better smearing for quark propagators so that the lowest state saturates the correlation function at short times, and to significantly increase the statistics.

## 8. Conclusions

We have presented a method for measuring the spin asymmetry of the proton using the Adler-Bell-Jackiw anomaly. Unfortunately, this method fails in the quenched approximation due to the $\eta'$ pole at $q^2 = 0$. We discuss the calculation of the topological charge density using the simplest field-theoretic method and explore numerical methods for obtaining a signal for the matrix element of $F\tilde{F}$ inside a proton using an existing set of quenched lattices. We do not find a reliable statistical signal and expect that future calculations using unquenched lattices will be needed to explore alternate methods to improve the signal.

## Acknowledgements

We acknowledge the support of DOE in the form of a Grand Challenge award of computer time at NERSC at Livermore. A significant fraction of the calculation was also done at the Pittsburgh Supercomputer Center and we are grateful to R. Roskies for his support. We also thank Aneesh Manohar and Steve Sharpe for discussions.

# References

[1] J. Ashman, *et al.*, *Phys. Lett.* **206B** (1988) 364; *Nucl. Phys.* **B328** (1989) 1.

[2] J. Kodaira, *Nucl. Phys.* **B165** (1979) 129 ;
D. Kaplan and A. Manohar, *Nucl. Phys.* **B310** (1988) 527

[3] Review of Particle Properties, *Phys. Rev.* **D45** (1992) VIII.9.

[4] R. Jaffe and A. Manohar, *Nucl. Phys.* **B337** (1990) 509.

[5] P. L. Anthony, *et al.*, *Phys. Rev. Lett.* **71** (1993) 959.

[6] B. Adeva, *et al.*, *Phys. Lett.* **302B** (1993) 533 ;
Preliminary results presented by G. Mallot at Yale University Conference *Internal Spin Structure of the Proton*, Jan 1994.

[7] J. Ellis and M. Karliner, *Phys. Lett.* **313B** (1993) 131.

[8] S. Y. Hseuh, *et al.*, *Phys. Rev.* **D38** (1988) 2056.

[9] J.E. Mandula and M.C. Ogilvie, *Phys. Lett.* **312B** (1993) 327.

[10] See J.E. Mandula, in Proceedings of ARC Conference on Polarization Dynamics, Trieste (1992) (hep-lat/9206027) for a preliminary discussion

[11] J. Smit and J. Vink, *Nucl. Phys.* **B284** (1987) 234.

[12] J. Mandula, J. Govaerts and G. Zweig, *Nucl. Phys.* **B228** (1983) 109.

[13] M. Campostrini, *et al.*, *Phys. Lett.* **252B** (1990) 436.

[14] M. Campostrini, *et al.*, *Nucl. Phys.* **B329** (1990) 683.

[15] A. Di Giacomo and A. Vicari, *Phys. Lett.* **275B** (1992) 429.

[16] M. Campostrini, *et al.*, *Phys. Lett.* **252B** (1990) 436.

[17] P. Lepage and P. Mackenzie, *Phys. Rev.* **D48** (1993) 2250.

[18] D. J. R. Pugh and M. Teper, *Phys. Lett.* **218B** (1989) 326 ;
M. Göckler, A. Kronfeld, M. Laursen, G. Schierholz and U. Wiese, *Phys. Lett.* **233B** (1989) 192

[19] G. Schierholz, *"LATTICE 92"*, Proceedings of the International Symposium on Lattice Field Theory, Amsterdam, The Netherlands, 1992, Eds. J. Smit *et al.*, *Nucl. Phys.* **B** (*Proc. Suppl.*) **30**, (1993)

[20] D. Daniel, R. Gupta, G. Kilcup, A. Patel, S. Sharpe, *Phys. Rev.* **D46** (1992) 3003.

|          | $T_{F\tilde{F}}=1$ | $T_{F\tilde{F}}=2$ | $T_{F\tilde{F}}=3$ | $T_{F\tilde{F}}=4$ | $T_{F\tilde{F}}=5$ | $T_{F\tilde{F}}=6$ | $T_{F\tilde{F}}=7$ | $T_{F\tilde{F}}=8$ |
|----------|--------|---------|---------|---------|---------|---------|---------|---------|
| $T_P=4$ | 4(07)  | $-8(07)$  | 0(07)   |         |         |         |         |         |
| $T_P=4$ | 5(08)  | $-4(06)$  | $-6(08)$  |         |         |         |         |         |
| $T_P=4$ | 6(06)  | 3(05)   | 13(05)  |         |         |         |         |         |
| $T_P=5$ | 8(09)  | $-17(11)$ | $-8(10)$  | $-11(09)$ |         |         |         |         |
| $T_P=5$ | 0(13)  | $-7(14)$  | 3(14)   | $-7(13)$  |         |         |         |         |
| $T_P=5$ | 7(09)  | 1(08)   | 15(08)  | 4(08)   |         |         |         |         |
| $T_P=6$ | 20(14) | $-23(15)$ | $-14(14)$ | $-16(12)$ | 14(12)  |         |         |         |
| $T_P=6$ | 1(20)  | $-19(25)$ | 30(24)  | $-18(16)$ | $-43(19)$ |         |         |         |
| $T_P=6$ | 0(12)  | $-1(13)$  | 17(13)  | 13(11)  | $-3(11)$  |         |         |         |
| $T_P=7$ | 28(22) | $-28(22)$ | $-23(19)$ | $-21(17)$ | 23(18)  | 30(26)  |         |         |
| $T_P=7$ | 28(30) | $-20(41)$ | 56(46)  | $-8(27)$  | $-66(36)$ | $-93(41)$ |         |         |
| $T_P=7$ | $-2(16)$ | 2(17)   | 19(18)  | 22(14)  | $-14(19)$ | 15(15)  |         |         |
| $T_P=8$ | 36(29) | $-29(29)$ | $-24(24)$ | $-17(29)$ | 28(27)  | 49(37)  | 40(36)  |         |
| $T_P=8$ | 44(40) | $-6(59)$  | 63(67)  | 1(50)   | $-96(55)$ | $-207(75)$ | $-47(78)$ |         |
| $T_P=8$ | $-1(19)$ | 13(24)  | 20(23)  | 31(19)  | $-13(25)$ | 26(23)  | $-6(25)$  |         |
| $T_P=9$ | 52(35) | $-15(39)$ | $-25(33)$ | $-9(41)$  | 26(36)  | 61(43)  | 50(50)  | $-22(40)$ |
| $T_P=9$ | 48(73) | 22(94)  | 38(94)  | 14(93)  | $-134(95)$ | $-305(118)$ | $-95(102)$ | $-13(99)$ |
| $T_P=9$ | $-6(27)$ | 32(31)  | 39(28)  | 39(24)  | $-21(29)$ | 34(32)  | $-7(30)$  | $-21(25)$ |

Table 1: The data for the matrix element as a function of $T_{F\tilde{F}}$ (time-slice of operator insertion) and $T_P$ (sink time-slice of the proton) at $\kappa = 0.154$. The source time-slice of the proton is held fixed at $T = 0$. The three numbers at each value of $T_P$ and $T_{F\tilde{F}}$ are for the three cases of momentum transfer, i.e. $\vec{q} = (2,0,0)$, $\vec{q} = (4,0,0)$, and $\vec{q} = (1,1,0)$ respectively. All numbers should be multiplied by $10^{-2}\alpha_s$ to get $\Delta\Sigma$.

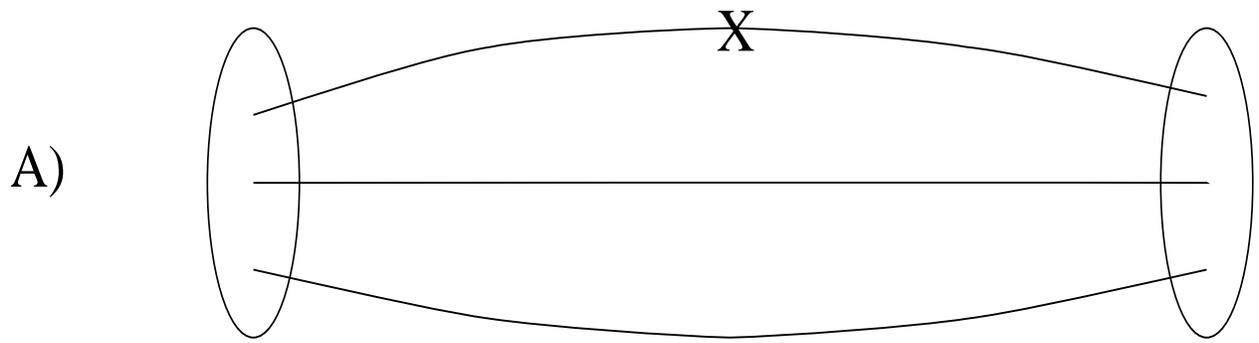

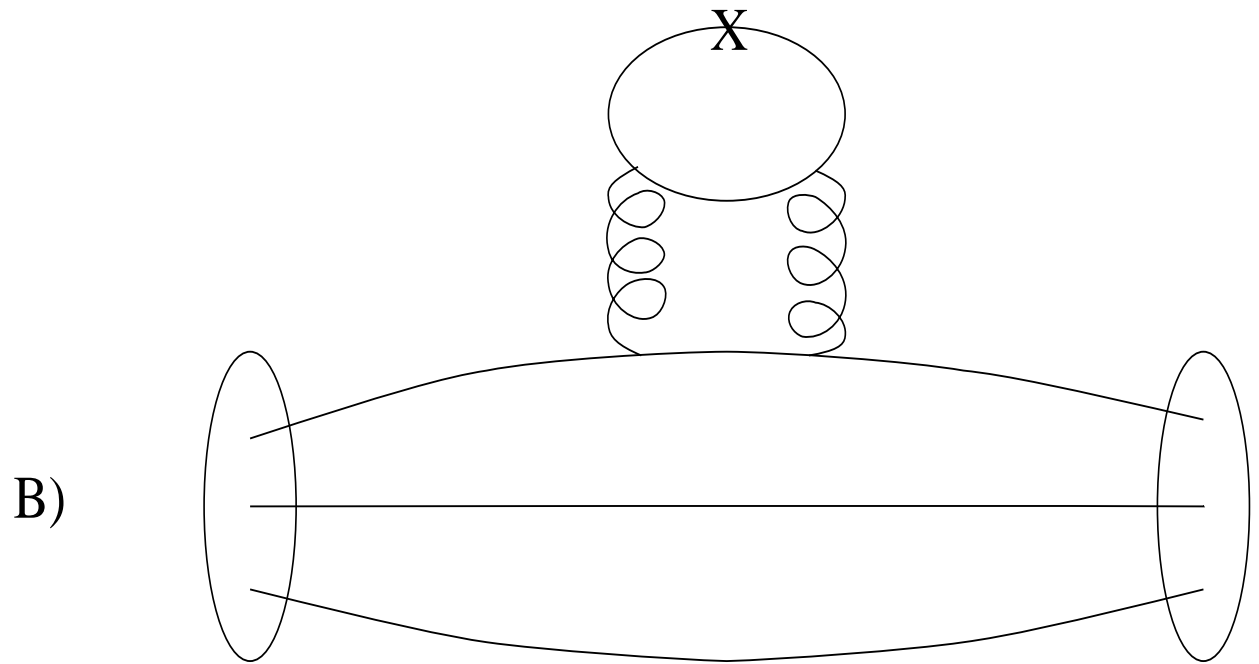

Fig. 1: The connected (a) and disconnected diagram (b) that contribute to the matrix element of the singlet axial current in a proton.

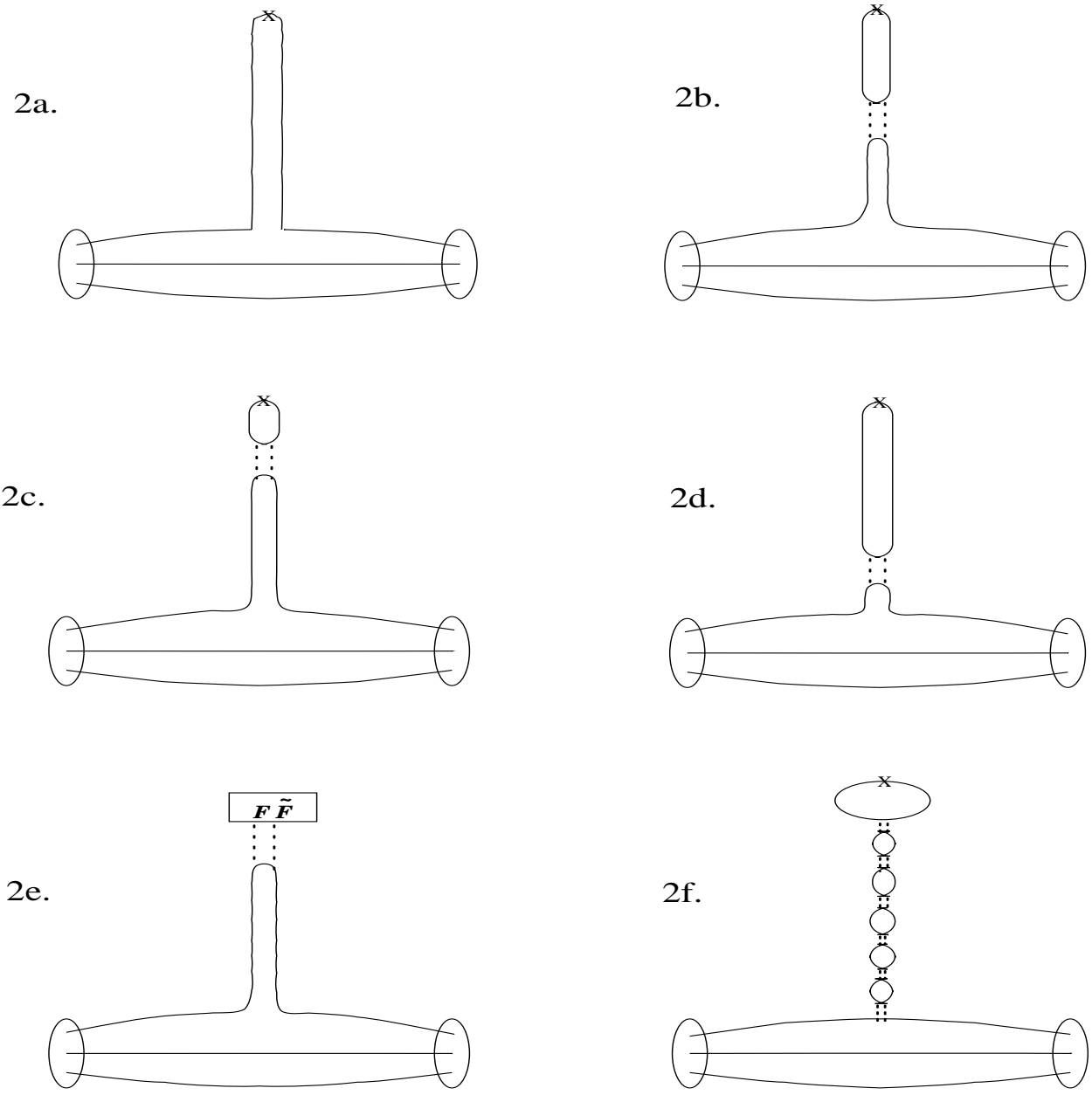

Fig. 2: Diagrams showing the contribution of $\eta'$ poles. The symbol $X$ denotes the insertion of the current and the dotted lines represent gluon exchange. Fig. 2a shows the single pole in the connected diagram, Figs. 2b-e show double and single poles in the disconnected diagram. In the full theory Figs. 2a-d are part of the same single pole shown in Fig. 2f.

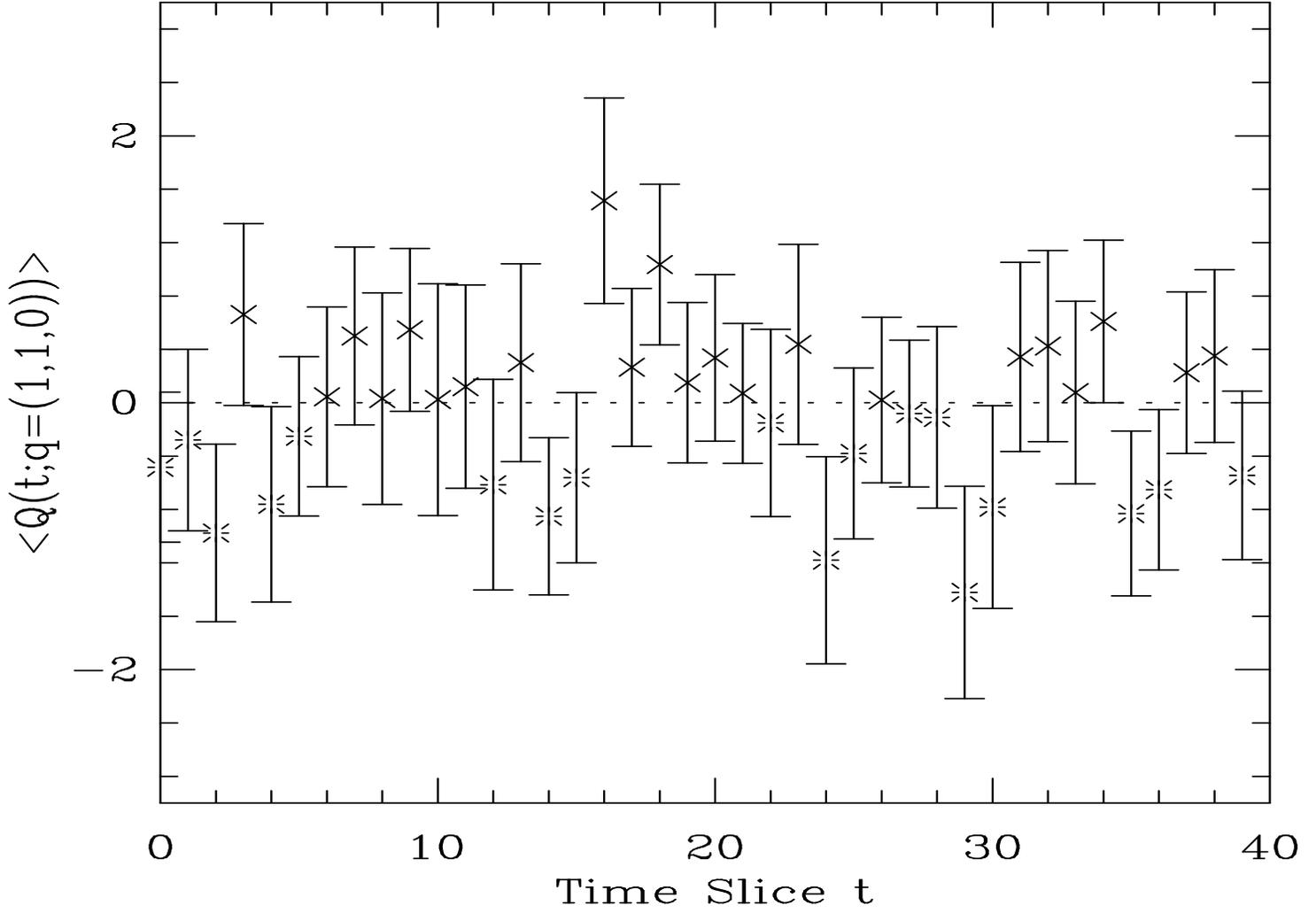

Fig. 3: The expectation value of Tr$F\tilde{F}$ as a function of time slice for $\vec{q} = (1,1,0)$. Different symbols have been used to help distinguish between the positive and negative values.

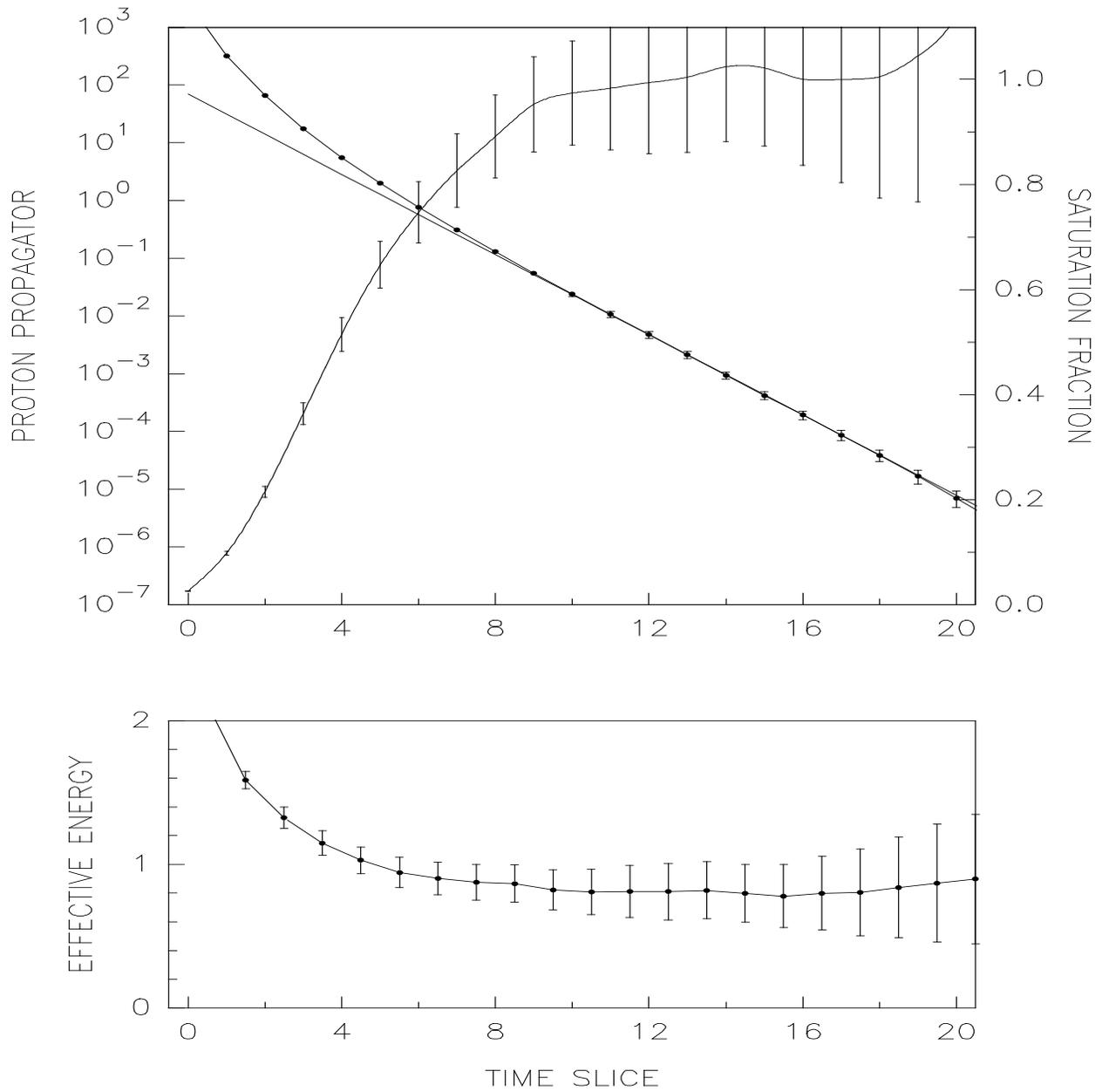

Fig. 4: The proton correlator and the single mass fit. We also show the saturation by the lowest state and the effective mass plot as a function of the separation t. The data are for $\vec{p} = (1,0,0)$.